# Quantum Scattering Model of Dual-Parallel Mach–Zehnder Electro-Optic Modulators: General Formalism, Impairments, and Applications to Frequency-Bin Photonic Quantum States

MAR YUSTE,[1] CARLOS R. FERNÁNDEZ-POUSA,[2] JAVIER FRAILE-PELÁEZ,[3] AND JOSÉ CAPMANY[1,*]

[1]*Universitat Politècnica de València, Valencia, Spain*
[2]*Universidad Miguel Hernández de Elche, 03202 Elche, Spain*
[3]*Universidad de Vigo, Vigo, Spain*
**jcapmany@iteam.upv.es*

**Abstract:** We present a quantum-mechanical model of dual-parallel Mach–Zehnder electro-optic modulators (DP-MZMs), derived using the quantum scattering formalism [1,2]. We obtain the complete, unrestricted transformation of single-photon and coherent input states through the DP-MZM architecture, including independent radio-frequency drive conditions on each internal modulator, and recover the symmetric push-pull operating point as an exact special case. We numerically evaluate the resulting frequency-comb structure, quantify the device sensitivity to bias error, modulation-index imbalance, and RF noise, and use the model to design a DP-MZM-based frequency-bin qubit source, reporting its fidelity as a function of these impairments.

## 1. Introduction

Electro-optic modulators are fundamental to modern photonic systems, and their quantum-mechanical description is required whenever optical signals are represented by quantum states rather than classical fields. Quantum models of electro-optic phase [1] and amplitude [2] modulators have been developed in the past using scattering formalism, describing modulation in terms of optical field operators. Many practical transmitters, however, are based on dual-parallel Mach–Zehnder modulators (DP-MZMs), whose nested interferometric architecture enables independent in-phase/quadrature control and has made them the standard device for classical single-sideband (SSB) and vector (I/Q) modulation in coherent communications and microwave photonics [3].

The classical description [3] is insufficient once a DP-MZM acts on non-classical light: a classical transfer function specifies only the mean field amplitude and captures neither photon statistics, vacuum fluctuations at unused beamsplitter ports, nor the operator non-commutativity governing single-photon detection probabilities and multi-photon interference. This is already practically relevant, since dual-parallel modulators are used experimentally as single-photon frequency shifters, exploiting the SSB mechanism of [3] to improve Hong–Ou–Mandel interference visibility between otherwise distinguishable photons [4]; predicting such experiments requires tracking the quantum state itself, not its classical power spectrum.

A second motivation is the growth of frequency-bin encoding for photonic quantum information: quantum frequency combs in integrated microresonators can host high-dimensional entangled states across many bins [5], and frequency-bin-entangled photon pairs with tens of modes have been demonstrated [6]. Electro-optic frequency beamsplitters and tritters have already been proposed for high-fidelity frequency-bin manipulation [7]. A DP-MZM is functionally a programmable frequency-domain beamsplitter, so a model predicting its action bin-by-bin directly supports the area of frequency-bin quantum information processing generally [8].

A third motivation is practical: quadrature bias drifts, I/Q drive amplitudes are never perfectly matched, and beamsplitters do not split light in an exact 3-dB ratio. Classical coherent communications address this with dedicated bias-control techniques [9], confirming these are first-order imperfections. In a quantum source, the same imperfections impair state fidelity, requiring an operator-level model rather than a classical bias-control analysis. A general quantum model of the DP-MZM is therefore needed to predict multi-photon interference in frequency-shifting experiments [4], design frequency-bin qubit sources compatible with quantum frequency-comb platforms [5,6], translate device imperfections into fidelity budgets [9], and connect DP-MZM parameters to downstream quantum-information performance, including quantum key distribution source characterization [10].

In this paper we develop, from first principles using the quantum scattering formalism of [1,2] and outlined in Section 2, the complete quantum model of the DP-MZM. We derive the general transformation for single-photon and coherent-state inputs without restricting beamsplitter ratios, bias points, or RF drive to be identical across the I and Q branches (Sections 3–4), recovering the symmetric push-pull operating point as an exact special case. We then evaluate the resulting frequency-comb structure (Section 5), quantify how extinction ratio, bias error, and I/Q imbalance degrade fidelity (Section 6), and apply the model to the design of a frequency-bin qubit source (Section 7).

## 2. DP-MZM Device Representation and Operator Formalism

We consider the DP-MZM device shown in Fig. 1. From left to right, it is composed of an input Y-branch beamsplitter $BS_i$ that opens two arms of an external Mach–Zehnder interferometer closed by a Y-branch output beamsplitter $BS_o$. In the upper, or *in-phase*, arm there is an internal MZI composed of an input $BS_1^{(I)}$ and an output $BS_2^{(I)}$ Y-branch beamsplitter that close two arms (upper +, lower −), where phase modulators $PM_+^{(I)}$ and $PM_-^{(I)}$ are respectively inserted. Similarly, in the lower, or *quadrature*, arm there is an internal MZI composed of an input $BS_1^{(Q)}$ and an output $BS_2^{(Q)}$ Y-branch beamsplitter closing two arms, with phase modulators $PM_+^{(Q)}$ and $PM_-^{(Q)}$. A static phase shifter $\phi_m$ is included in the quadrature arm for biasing purposes.

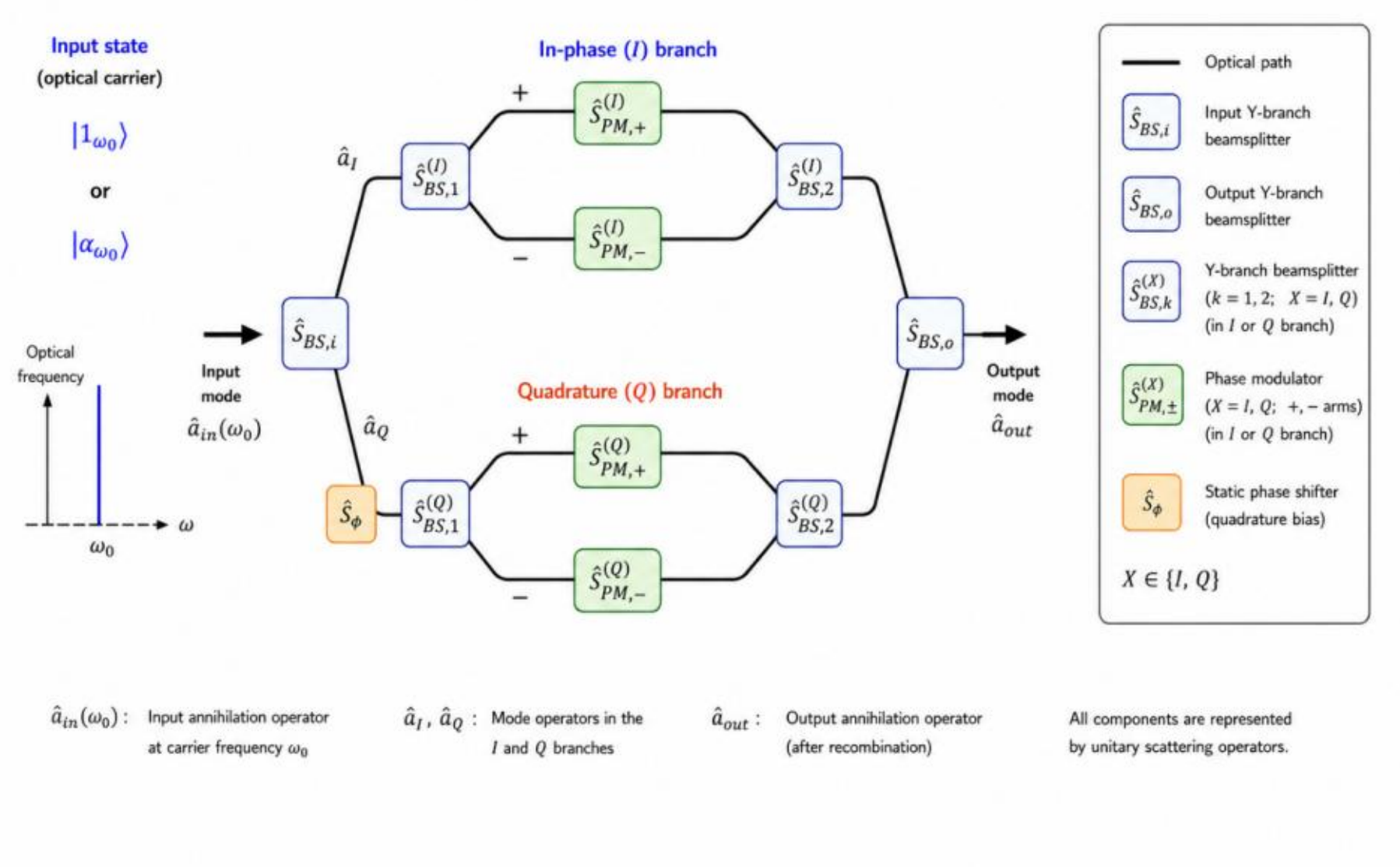

Fig. 1. Layout of the DP-MZM device, showing the input/output external beamsplitters, the in-phase and quadrature internal MZIs with their branch phase modulators, and the quadrature bias phase shifter, together with the unitary operators representing each component.

Each beamsplitter is represented by a unitary scattering operator acting on the pair of field-mode creation operators at its two input ports [11]:

$$\hat{S}_{BS}\begin{pmatrix}\hat{a}_1\\ \hat{a}_2\end{pmatrix}\hat{S}_{BS}^{\dagger}=\begin{pmatrix}t & r\\ -r & t\end{pmatrix}\begin{pmatrix}\hat{a}_1\\ \hat{a}_2\end{pmatrix} \tag{1}$$

where $t$ and $r$ are the (generally complex) field transmission and reflection coefficients, subject to the unitarity constraint $|t|^2+|r|^2 = 1$ and the usual $\pi/2$ relative phase difference between the bar and cross ports. Each splitting Y-branch is understood to have a fictitious input port fed by a vacuum operator, and each combining Y-branch a fictitious output port producing a radiated mode of no interest to the model; both are needed only to preserve unitarity of the overall transformation and are discarded once the physical output mode is identified.

The overall action of the DP-MZM is represented by a unitary operator $\hat{S}_{DP}$ built from the composition, in the order shown in Fig. 1, of the input splitter, the four branch phase modulators, the two pairs of internal combiners/splitters, the quadrature bias phase shifter, and the output combiner. The quantum model amounts to computing the transformation $\hat{S}_{DP}\ \hat{A}\ \hat{S}_{DP}^{\dagger}$ of the relevant operator $\hat{A}$ representing the quantum state of interest: the creation operator

$$\hat{a}_{n_o}^{\dagger}$$

for a single photon of frequency $\omega_o$, or the displacement operator $\hat{D}(\alpha)$ for a coherent state at frequency $\omega_o$, following the standard quantum-optical representation of these states [11]. Sections 3 and 4 develop these two cases in full generality, with the complete stage-by-stage operator algebra given in Appendices A and B, respectively; Sections 3.3 and 4.3 specialize this general result to the practically important symmetric, push-pull operating point.

## 3. General Quantum Model for Single-Photon Input States

The full stage-by-stage operator algebra summarized in this section is given in complete detail in Appendix A; here we present the structure of the derivation and its two key results, Eq. (5) and its special case Eq. (6).

### *3.1 Stage-by-stage transformation*

We proceed from left to right through the device, propagating the single-photon creation operator $\hat{a}_{no}^{\dagger} \otimes 1$ (with the second factor labelling the vacuum in the fictitious input mode) through each stage. The input beamsplitter $\hat{S}_{BS,i}$ splits the mode into the in-phase and quadrature arms; because each arm subsequently opens into its own internal MZI, the two-mode Hilbert space at the input becomes a four-mode product space (two in the in-phase arm, two in the quadrature arm) once the internal input beamsplitters $\hat{S}_1^{(I,Q)}$ act. Each of the four resulting modes then acquires the frequency-comb sideband structure imposed by its respective phase modulator, described in general by the transformation

$$\hat{S}_{PM,+}^{(I)}\,\hat{a}_{n_o}^{\dagger}\,\hat{S}_{PM,+}^{(I)\dagger}=\sum_{q=1}^{\infty}C_{q,+}^{(I)}\left(q_o^{(I)}\right)\hat{a}_{qN_{o,+}^{(I)}-r_{o,+}^{(I)}}^{\dagger} \tag{2}$$

where the coefficients $C_{q,+}^{(I)}\left(q_o^{(I)}\right)$ follow the quantum phase-modulation result of ref. [1]:

$$C_{q,+}^{(I)}\left(q_o^{(I)}\right)=e^{iq\theta_{b,+}^{(I)}}\left(ie^{i\theta_{b,+}^{(I)}}\right)^{q-q_o^{(I)}}\left[J_{q-q_o^{(I)}}\left(m_+^{(I)}\right)-(-1)^{q_o^{(I)}}J_{q+q_o^{(I)}}\left(m_+^{(I)}\right)\right] \tag{3}$$

with $J_x$ the Bessel function of the first kind, $m_+^{(I)}$ the modulation index and $\theta_{b,+}^{(I)}$ the bias phase of the upper in-phase modulator, and where $\omega_o = 2\pi n_o c/L$, $n_o = q_o^{(I)}\,N_o^{(I)} - r_o^{(I)}$ with $0 \leq r_o^{(I)} < N_o^{(I)}$ being the discrete comb decomposition of the input carrier frequency relative to the RF drive frequency $\Omega = 2\pi N_o^{(I)}c/L$ (L a formal quantization length). In the regime of practical interest, where the number of significantly populated sidebands is much smaller than $N_o^{(I)}$, this reduces to the more transparent form

$$\hat{S}_{PM,+}^{(I)}\,\hat{a}_{n_o}^{\dagger}\,\hat{S}_{PM,+}^{(I)\dagger} \;\approx\; e^{iq_o\theta_{b,+}}\sum_{s=1-q_o}^{\infty}\left(ie^{i\theta_{b,+}}\right)^{s}J_s(m_+)\hat{a}_{n_o+sN_o}^{\dagger} \tag{4}$$

i.e., an input carrier line generates sidebands at integer multiples $s$ of the RF frequency, each weighted by a Bessel coefficient of the modulation index and a phase set by the bias and drive phase — the standard quantum-optical signature of phase modulation. Analogous expressions hold for the lower in-phase modulator ($\mathrm{PM}_{-}^{(I)}$) and for both quadrature-branch modulators ($\mathrm{PM}_{+}^{(Q)}$, $\mathrm{PM}_{-}^{(Q)}$), each with its own, in general independent, modulation index and bias.

The two internal-arm modes are then recombined by the output internal beamsplitters $\hat{S}_2^{(I,Q)}$, which conserve the four-mode dimensionality, after which the quadrature arm passes through the static bias phase shifter modeled as multiplication by $e^{i\phi_P}$, before the final output beamsplitter $\hat{S}_{BS,o}$ recombines the in-phase and quadrature arms into the physical output mode plus a fictitious radiated mode, which is discarded. All in all, the model reduces to a one-input, one-output-port configuration, illustrated in the frequency domain in Fig. 2.

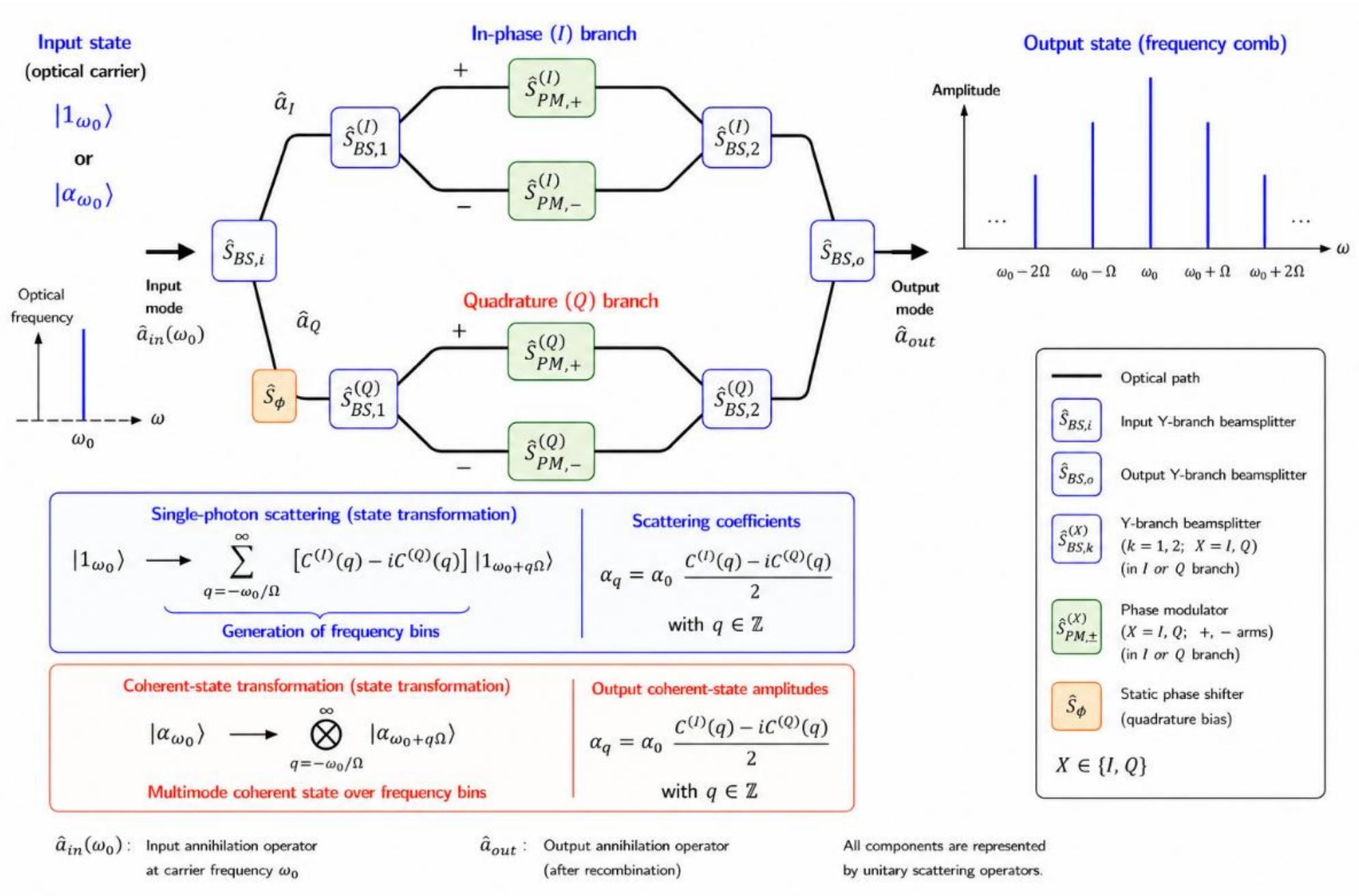

Fig. 2. Frequency-domain representation of quantum scattering in the DP-MZM. An input carrier state at frequency $\omega_o$ is transformed into a superposition of sidebands at $\omega_o + q\Omega$, forming a frequency-comb structure whose amplitudes are set by the coherent interference of the in-phase (I) and quadrature (Q) branches. Single-photon and coherent-state transformations are illustrated for the two input cases treated in Sections 3 and 4.

Figure 3 illustrates this comb structure numerically for a representative single-photon input, showing how sideband power redistributes as the modulation index increases.

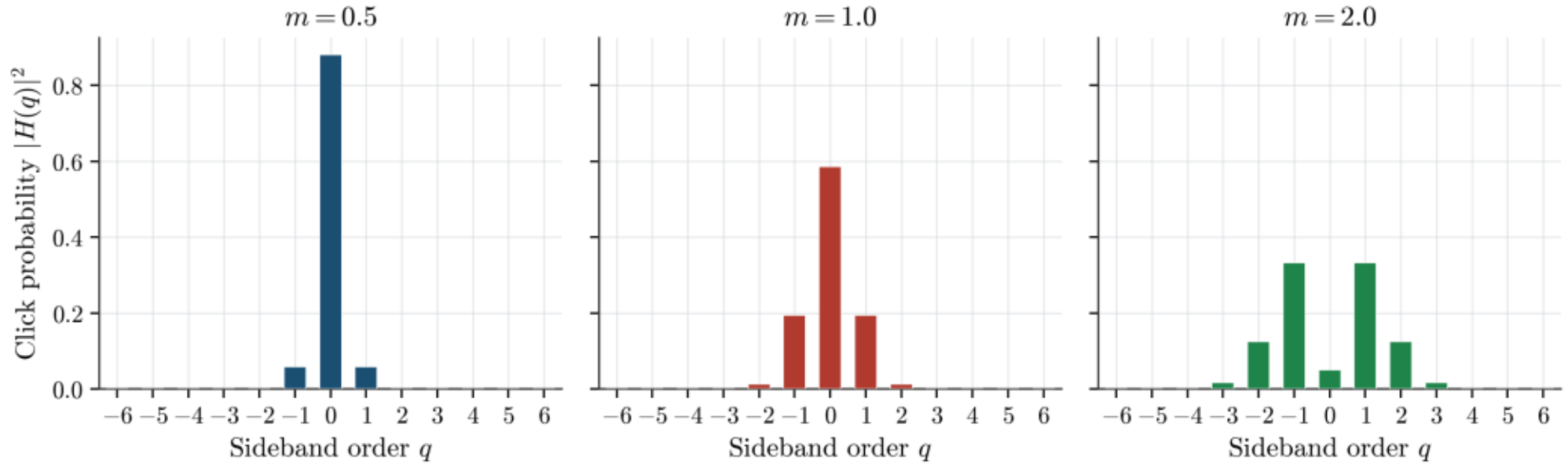

Fig. 3 (numerical illustration). Normalized single-photon sideband click probability $|H(q)|^2$ vs. sideband order q for three modulation indices, ideal push-pull operation, $\phi_m = \pi/2$, balanced 3-dB splitters.

### *3.2 General result*

Collecting the four branch contributions and carrying the beamsplitter coefficients through explicitly (rather than assuming 3-dB, balanced devices from the outset) gives the general, unrestricted single-photon transformation (full derivation is given in Appendix A.1–A.5):

$$\hat{S}_{DP}\,\hat{a}^{\dagger}_{n_o}\,\hat{S}^{\dagger}_{DP} = \sum_{q=1}^{\infty}\Big[t_o t_{I,2} t_{I,1} t_i\, C^{(I)}_{q,+}\big(q^{(I)}_o\big)\hat{a}^{\dagger}_{\dots} - t_o r_{I,2} r_{I,1} t_i\, C^{(I)}_{q,-}\big(q^{(I)}_o\big)\hat{a}^{\dagger}_{\dots} - -e^{i\phi_P} r_o t_{Q,2} t_{Q,1} r_i\, C^{(Q)}_{q,+}\big(q^{(Q)}_o\big)\hat{a}^{\dagger}_{\dots} + e^{i\phi_P} r_o r_{Q,2} r_{Q,1} r_i\, C^{(Q)}_{q,-}\big(q^{(Q)}_o\big)\hat{a}^{\dagger}_{\dots}\Big] \quad (5)$$

Equation (5) is the central result of this section: it holds for arbitrary beamsplitter power-splitting ratios on all six beamsplitters in the device, for independent modulation indices and bias phases on each of the four branch modulators, for independent RF drive frequencies on the in-phase and quadrature branches, and for an arbitrary quadrature bias $\phi_P$. It therefore also captures, as limiting cases, every impairment considered in Section 6.

### *3.3 Special case: the symmetric push-pull operating point*

If both ports of the in-phase and quadrature modulators are driven at the same RF frequency and with $\phi_P = \pi/2$, all four sideband indices and comb parameters coincide ($N_o^{(I)} = N_o^{(Q)} = N_o$, $q_o^{(I)} = q_o^{(Q)} = q_o$, $r_o^{(I)} = r_o^{(Q)} = r_o$). If, further, all beamsplitters are balanced ($t = r = 1/\sqrt{2}$) and each internal MZM is operated in push-pull configuration, so that $C_{q,-}(q_o) = -C_{q,+}(q_o)$ on both branches, Eq. (5) transforms to

$$\hat{S}_{DP}\,\hat{a}^{\dagger}_{n_o}\,\hat{S}^{\dagger}_{DP} = \frac{1}{2}\sum_{q=1}^{\infty}\Big[C^{(I)}_{q,+}(q_o) - i\,C^{(Q)}_{q,+}(q_o)\Big]\,\hat{a}^{\dagger}_{qN_o - r_o} \quad (6)$$

This is the practically important symmetric, push-pull operating point. Equation (6) makes explicit that the ideal DP-MZM acts as a quantum frequency-scattering device implementing a complex, weighted sum of the in-phase and quadrature branch sideband spectra — the optical analogue of a classical I/Q modulator.

## 4. General Quantum Model for Coherent Input States

As in Section 3, the complete stage-by-stage derivation is given in Appendix B; here we summarize its logic and its two key results.

### *4.1 Action of the DP-MZM on the displacement operator*

For a coherent-state input we must compute the transformation $\hat{S}_{DP}$ ($\hat{D}_{no}(\alpha_{no})$ $\otimes$ 1) $\hat{S}_{DP}^{\dagger}$. Because every element of the DP-MZM is represented by a unitary operator that is linear in the mode creation/annihilation operators (beamsplitters and phase modulators alike), its action on a displacement operator reduces, via the Baker–Campbell–Hausdorff theorem, to the same linear combination of output modes found for the creation operator, but exponentiated into a product of displacement operators on those modes (derived in Appendix B.1–B.2):

$$\hat{S}\,\hat{D}_n(\alpha_n)\,\hat{S}^{\dagger} = \hat{D}_n(\alpha_n) \otimes \mathbb{1} \quad \Rightarrow \quad \hat{S}\,\hat{a}_n^{\dagger}\,\hat{S}^{\dagger} = \sum_k c_k\ \hat{a}_k^{\dagger} \ \Leftrightarrow\ \alpha_k = \alpha_n\, c_k \quad (7)$$

In other words, once the single-photon (creation-operator) transformation of Section 3 is known for a given DP-MZM configuration, the coherent-state transformation follows immediately by replacing each output creation operator with a displacement operator carrying the same complex weight, multiplied by the input amplitude $\alpha_{no}$.

### *4.2 General result*

Carrying out this replacement through the complete device, in exact parallelism with Eq. (5), gives (see Appendix B.3–B.7)

$$\hat{S}_{DP}\big(\hat{D}_{n_o}\big(\alpha_{n_o}\big) \otimes 1\big)\hat{S}_{DP}^{\dagger} = \hat{S}_{DP}\hat{D}_{n_o}\big(\alpha_{n_o}\big)\hat{S}_{DP}^{\dagger} \otimes 1 \ \propto\ \prod_{q=1}^{\infty} \hat{D}_{qN_o-r_o}\left(\alpha_{qN_o-r_o}\right) \quad (8)$$

where each output-mode amplitude $\alpha_{qN_o-r_o}$ is obtained from the input amplitude $\alpha_{no}$ weighted by the same combination of beamsplitter coefficients and Bessel sideband coefficients $C_{q,\pm}^{(I,Q)}(q_o)$ that appears in Eq. (5) for the single-photon case.

### *4.3 Special case: the symmetric push-pull operating point*

Under the same simplifying assumptions as Section 3.3 (common RF frequency, $\phi_P = \pi/2$, balanced beamsplitters, push-pull operation), the general coherent-state result reduces to

$$\alpha_{qN_o-r_o} = \frac{\alpha_{n_o}}{2}\left[C_{q,+}^{(I)}(q_o) - i\, C_{q,+}^{(Q)}(q_o)\right] \quad (9)$$

This is the coherent-state analogue of Eq. (6), and shows that in this ideal limit the sideband amplitude weighting is identical for the single-photon and coherent-state cases — as expected, since the DP-MZM is a linear, passive device and the two cases differ only in which quantum state of light is being scattered by the same classical-field-equivalent transfer function.

### *4.4 Comparison of the two input cases*

It is instructive to display the single-photon click-probability spectrum and the coherent-state mean-photon-number spectrum side by side, since both are governed by the same underlying transfer function *H(q)* of Eq. (10) below (Section 5). Figure 4 shows this comparison at a representative operating point ($m$ = 1.2, ideal bias): the two spectra share the same normalized envelope shape, differing only in their physical interpretation — a per-photon detection probability in one case, a mean photon-number distribution in the other.

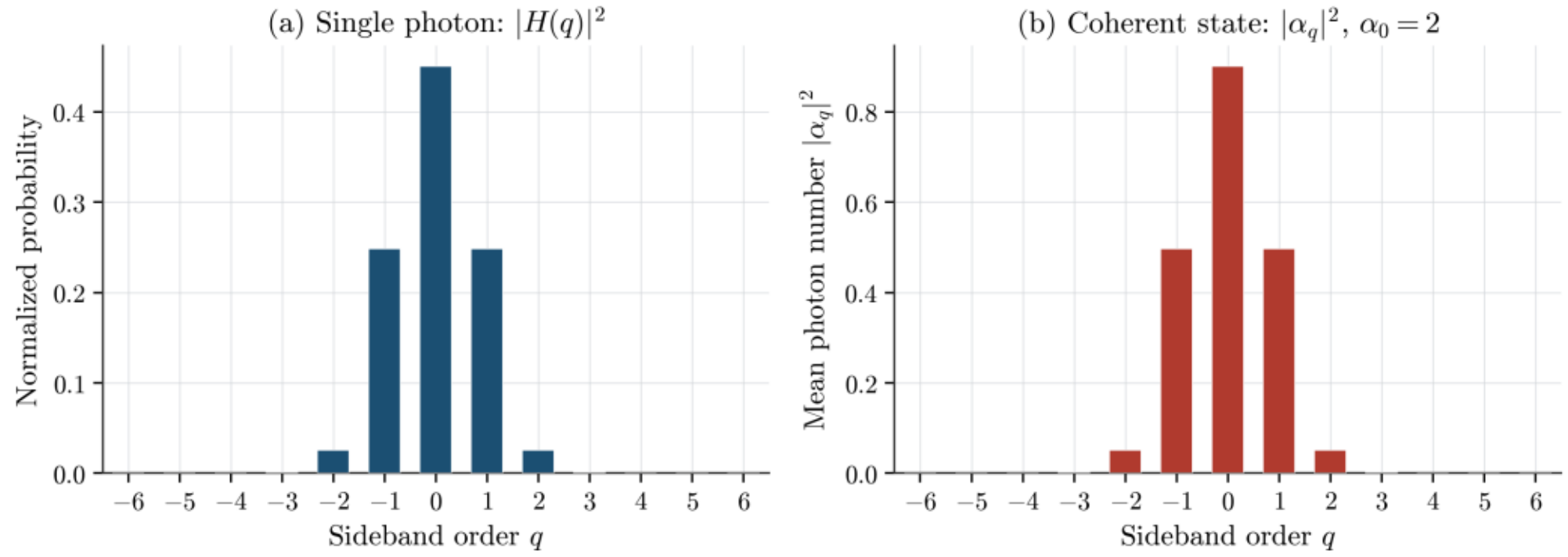

Fig. 4. Side-by-side comparison of (a) the single-photon click-probability spectrum and (b) the coherent-state mean-photon-number spectrum (input amplitude $\alpha_0 = 2$), for identical DP-MZM operating conditions (m = 1.2, ideal push-pull bias). The two envelopes coincide up to normalization, confirming Eq. (9).

## 5. Numerical Results: Frequency-Comb Structure

To visualize the sideband structure predicted by the model and to prepare the impairment and application analysis of Sections 6–7, we evaluate the ideal push-pull transfer function implicit in Eqs. (6) and (9),

$$H(q) \equiv \frac{1}{2}\left[C_{q,+}^{(I)}(q_o) - i\,C_{q,+}^{(Q)}(q_o)\right], \qquad C_{q,+}(q_o) \simeq J_q(m)\,e^{iq\theta_b} \tag{10}$$

where the approximation $C_{q,+}(q_o) \simeq J_q(m)\ e^{iq\theta_b}$ follows from Eq. (4) and is adopted throughout this section and Sections 6–7 for numerical evaluation; it retains the full Bessel-function sideband structure and the bias-dependent phase of the general model while normalizing away the device-length-dependent comb bookkeeping ($n_o$, $N_o$, $r_o$) of Eqs. (2)–(3), which is immaterial once a fixed carrier and RF frequency are chosen. Figures 2–4 below use the symmetric operating point $\theta_I = \theta_Q$ (matching Eq. (6)/(9)); Fig. 6, and all of Sections 6–7, use the RF-quadrature operating point $\theta_Q = \theta_I - \pi/2$ instead, for the reason explained just before Fig. 6. In both cases $\phi_P = \pi/2$ and beamsplitters are ideal (3-dB, balanced) unless a parameter is explicitly being swept.

Figure 3 (Section 3.1) already showed the single-photon comb for three modulation indices at a common RF frequency: as $m$ increases, sideband power spreads from the carrier (q = 0) into higher orders following the Bessel-function envelope, exactly as expected for phase-modulation-type sideband generation.

Figure 5 illustrates the general, non-degenerate result of Section 3.2 (Eq. (5)) in a regime that the simplified symmetric-point equations (Eq. (6)/(9)) cannot represent, since those assume a single common RF modulating frequency: here the in-phase and quadrature branches are driven at different RF frequencies $\Omega_Q \neq \Omega_I$. Because the two branches' sidebands no longer fall on a common frequency grid, the output spectrum splits into two interleaved combs whose relative spacing tracks the frequency ratio $\Omega_Q/\Omega_I$ — a direct consequence of the general model of Section 3.2 that has no counterpart in the simplified treatment.

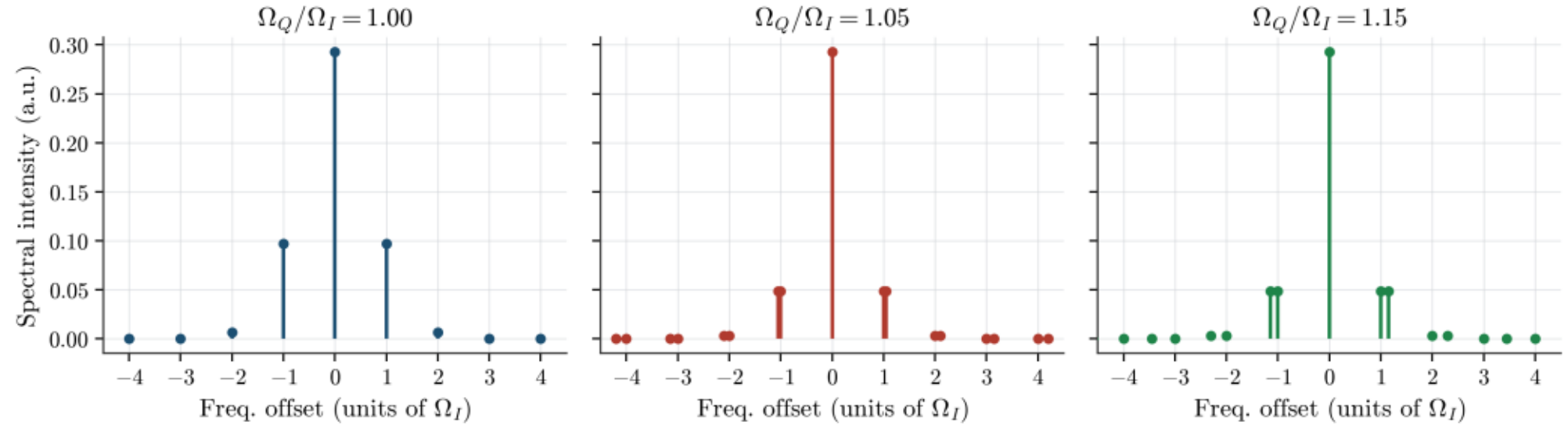

Fig. 5. Output spectral intensity for three values of the I/Q RF-frequency ratio Ω_Q/Ω_I, evaluated from the general (non-degenerate-frequency) transfer function of Section 3.2. Frequency axis in units of Ω_I; each stem is an exact comb line at its true frequency, with no numerical broadening.

Finally, Fig. 6 illustrates comb reshaping under bias error. This figure (and, as discussed in Section 6, the entire image-suppression analysis) requires an important qualification. At the symmetric operating point of Eq. (6)/(9) ($\theta_I = \theta_Q$, the 'ideal' case of Sections 3.3/4.3), the sideband amplitudes satisfy $|H(q)| = |H(-q)|$ identically, for any $\phi_P$ and any beamsplitter ratio — a direct consequence of $J_{-x}(m) = (-1)_x J_x(m)$. At that operating point there is therefore no well-defined 'intended vs. image sideband' to protect, and bias or splitter-ratio errors cannot, even in principle, create an imbalance between q and −q. Genuine single-sideband selectivity instead requires driving the I and Q branches in RF quadrature, $\theta_Q = \theta_I - \pi/2$ (a 90° phase offset between the two RF drive tones, on top of the π/2 optical combiner bias $\phi_m$) — the standard operating point for real single-sideband dual-parallel MZM transmitters. Figure 6 therefore adopts this RF-quadrature operating point, which is also the one used throughout Section 6 and Section 7.

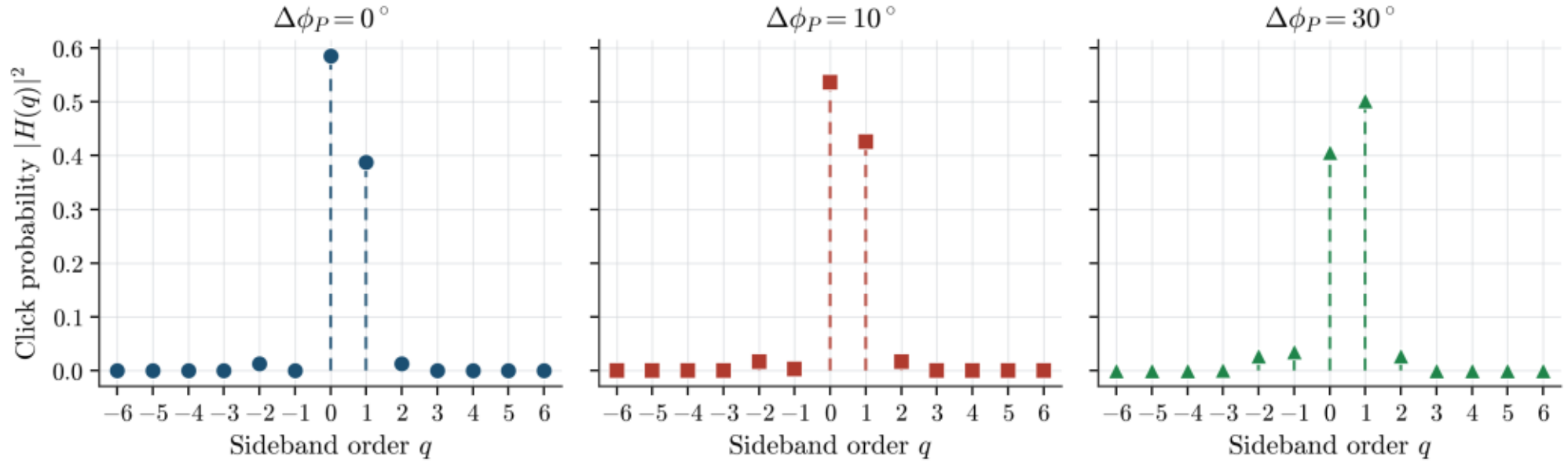

Fig. 6. Single-photon sideband spectrum at the RF-quadrature (single-sideband) operating point, for three values of the optical quadrature bias error $\Delta\phi_P$. At $\Delta\phi_P = 0$, odd-negative sidebands (q = −1, −3, …) are exactly nulled while q = 0, +1, +2, … survive; bias error partially restores the suppressed sidebands.

## 6. Impact of Practical Impairments

Loss, RF noise, finite extinction ratio, and I/Q imbalance are natural, practically important extensions of the ideal model. As established in Section 5, a meaningful image-sideband-suppression analysis requires the RF-quadrature (single-sideband) operating point $\theta_Q = \theta_I - \pi/2$, rather than the symmetric push-pull point of Eq. (6)/(9); all results in this section therefore use that operating point, with the modulation index at the nominal value $m_I = m_Q = 1$ unless stated otherwise. We first address four deterministic impairments captured directly by the general result of Section 3.2/4.2: quadrature-bias error, I/Q modulation-index imbalance, finite output-beamsplitter extinction ratio, and insertion loss (Sections 6.1–6.4). We then analyze genuine stochastic noise sources — RF phase noise, modulation-index jitter, and quadrature-bias drift — evaluated by Monte Carlo averaging over the same general model (Section 6.5).

### *6.1 Quadrature-bias error*

At the RF-quadrature operating point, deviating $\phi_P$ from $\pi/2$ by $\Delta\phi_P$ partially restores the sidebands that ideal single-sideband operation nulls, degrading the image-sideband suppression ratio,

$$\varepsilon_{\mathrm{dB}} = 10\log_{10}\left(\frac{P_q}{P_{-q}}\right) \tag{11}$$

evaluated between the q = +1 (intended) and q = −1 (image) sidebands. Figure 7 shows the resulting suppression as a function of $\Delta\phi_P$. At $\Delta\phi_P = 0$ the model gives essentially exact suppression. The suppression degrades to 20 dB at a bias error of Δϕ_P ≃ 11°, indicating that active bias control to within roughly ±10– 15° is required to preserve high spectral purity at this operating point.

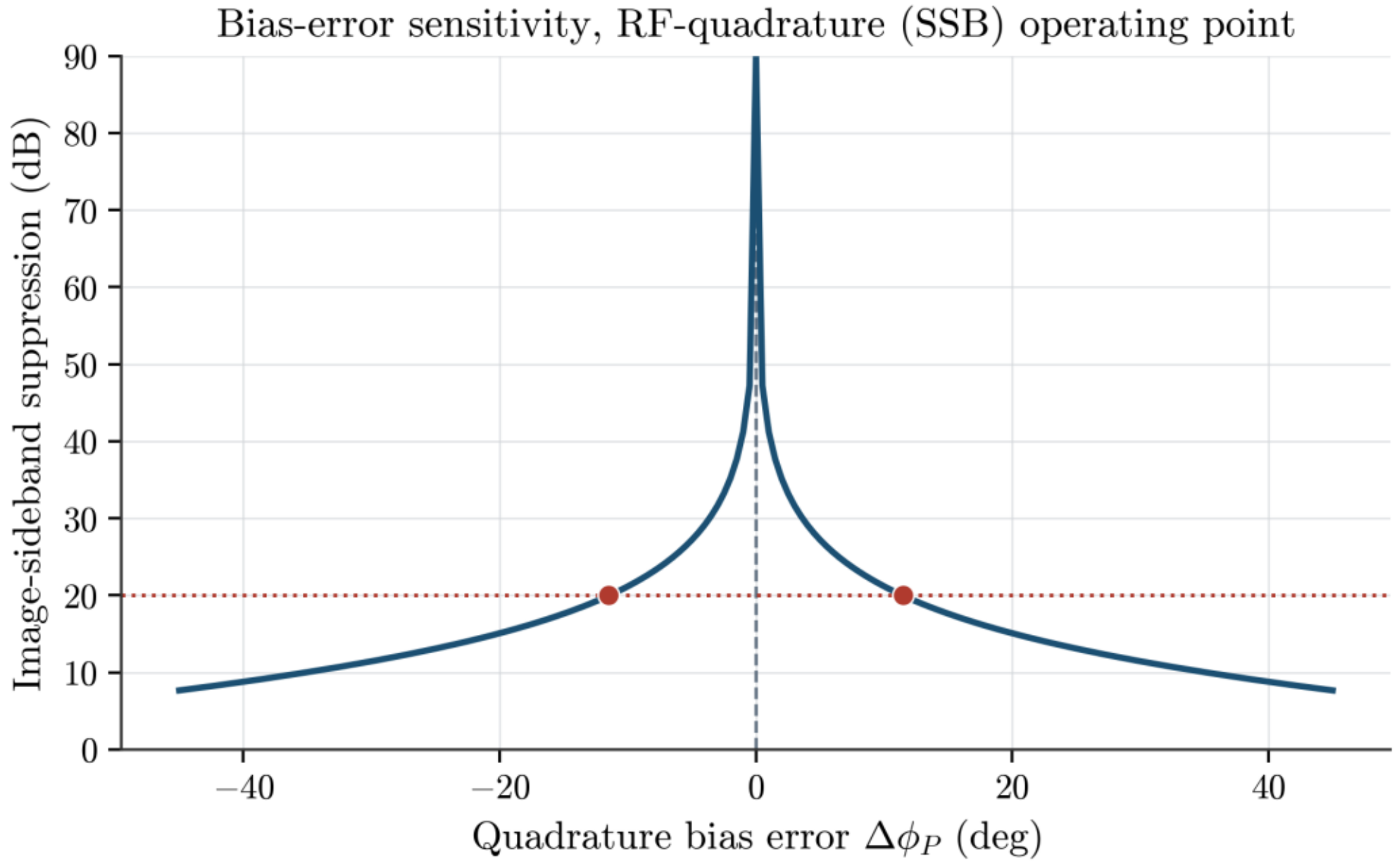

Fig. 7. Image-sideband suppression (q = +1 vs. q = −1) as a function of quadrature bias error Δϕ_P, at the RF-quadrature (single-sideband) operating point of Section 6.

### *6.2 I/Q modulation-index imbalance*

Independently varying the quadrature-branch modulation index $m_Q$ relative to the in-phase index $m_I$ (via Eq. (5), with $m_I$ fixed at the nominal value 1, RF-quadrature operating point) models a practical RF-driver amplitude mismatch between the two branches. We quantify its effect using the fidelity of the realized two-frequency-bin state to an ideal target qubit state,

$$F = |\langle\psi_{\mathrm{ideal}}|\psi_{\mathrm{realized}}\rangle|^2, \qquad |\psi_{\mathrm{ideal}}\rangle = \frac{1}{\sqrt{2}}(|q=0\rangle + |q=1\rangle) \tag{12}$$

where $|\psi_{\text{realized}}\rangle$ is the (normalized) DP-MZM output restricted to the q = 0, 1 subspace. At the perfectly matched point $m_Q/m_I$ = 1, the fidelity is F ≃ 0.846 — not unity, because, as shown in Section 7, this operating point (*m = 1*, away from the optimized $m$*) does not yet balance the *q = 0* and *q = 1* populations. Figure 8 shows that the fidelity varies smoothly between about 0.80 and 0.89 as $m_Q/m_I$ is swept over ±30%, i.e., I/Q index mismatch has a comparatively mild effect relative to the population-balance and fixed-phase considerations discussed in Section 7.

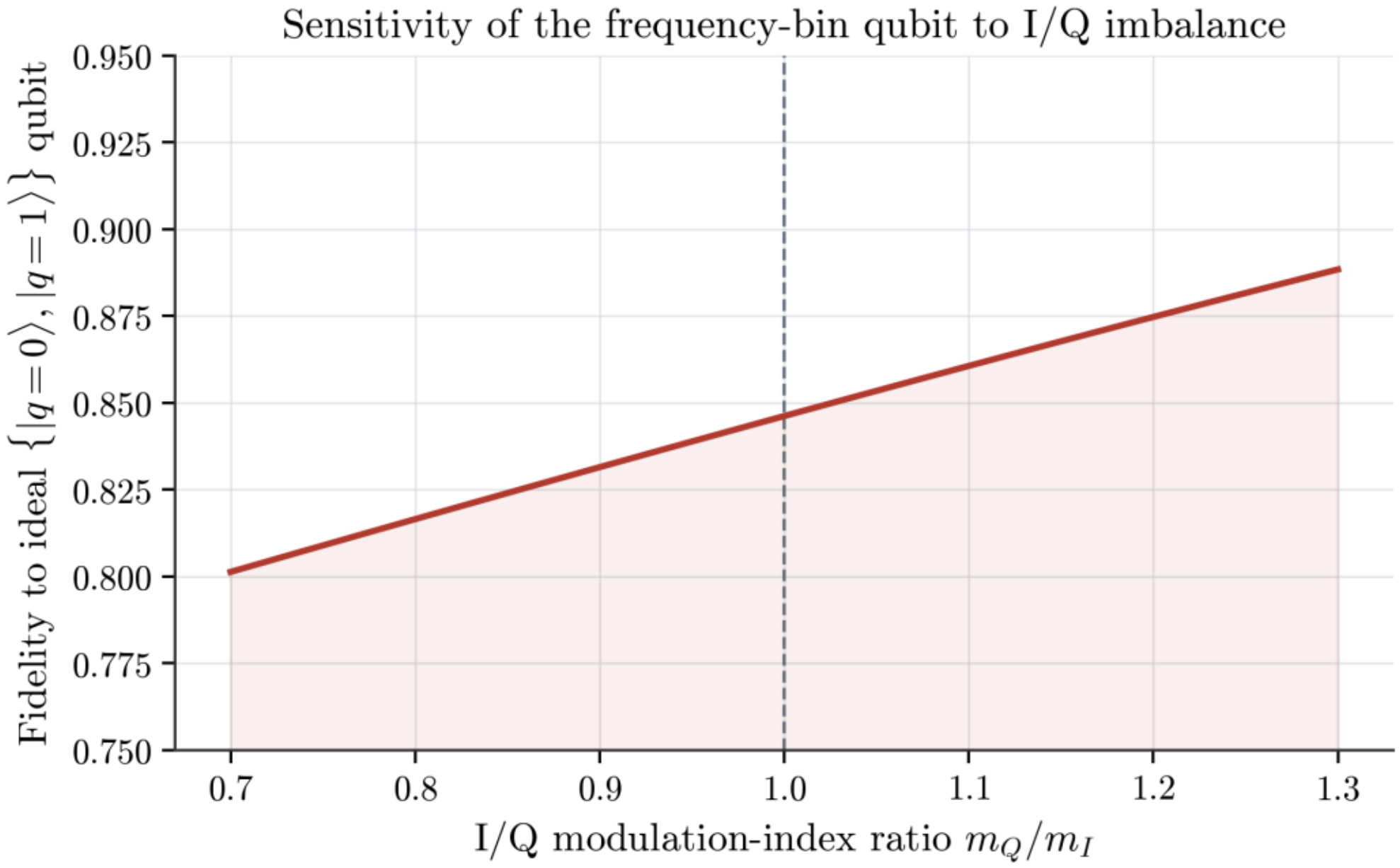

Fig. 8. Fidelity (Eq. (12)) of the realized two-frequency-bin state to the ideal target qubit as a function of the I/Q modulation-index ratio m_Q/m_I, at m_I = 1, RF-quadrature operating point.

### *6.3 Finite beamsplitter extinction ratio*

Relaxing the balanced-beamsplitter assumption at the output combiner ($t_o = (1/\sqrt{2})(1+\varepsilon)$, and $r_o = \sqrt{(1-t_o^2)}$, via Eq. (5), RF-quadrature operating point) models a fabrication-limited splitting ratio. Figure 9 shows the resulting leakage. At perfect balance (ε = 0) the model again gives essentially exact image suppression; a ±15% output-splitter power imbalance degrades the relative image leakage to approximately −17 dB, comparable in severity to a bias error of a few tens of degrees (Section 6.1). This underscores that both the internal and output Y-branch splitters should be fabricated close to their nominal 3-dB point for high spectral purity, in addition to the bias-control requirement of Section 6.1.

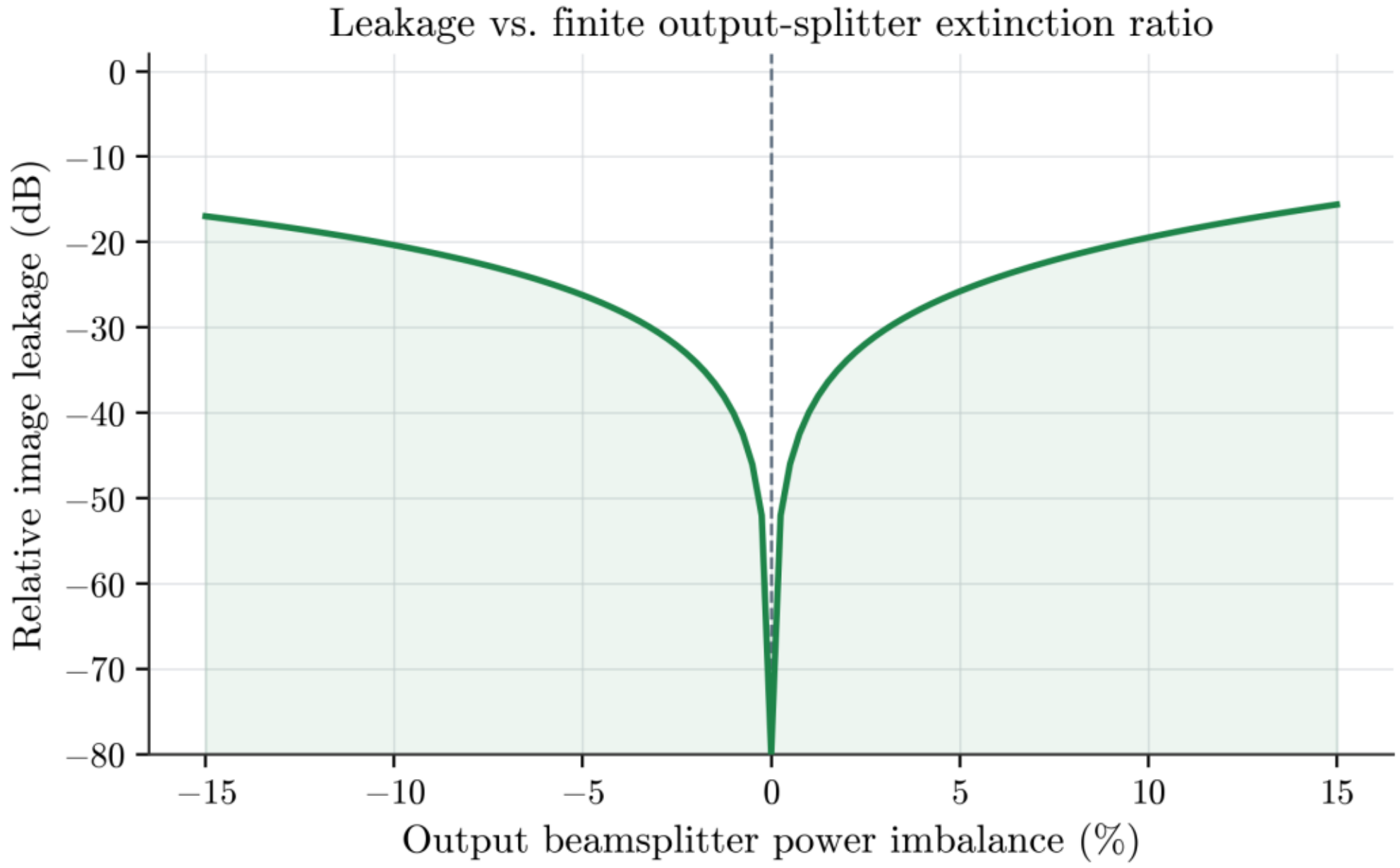

Fig. 9. Relative image-sideband leakage as a function of output-beamsplitter power-splitting imbalance, at the RF-quadrature (single-sideband) operating point of Section 6.

### 6.4 Insertion loss

Optical insertion loss at any stage of the device can be incorporated into the same operator formalism by coupling the relevant mode to an auxiliary vacuum (loss) mode through a fictitious beamsplitter of amplitude transmissivity $\sqrt{\eta}$,

$$\hat{a}^{\dagger}_{\mathrm{out}} = \sqrt{\eta}\,\hat{a}^{\dagger}_{\mathrm{in}} + \sqrt{1-\eta}\,\hat{v}^{\dagger}, \qquad \eta = 10^{-\mathrm{IL}/10} \tag{13}$$

where IL is the insertion loss in dB. For a coherent-state input this simply rescales the output amplitudes of Eq. (8) by $\sqrt{\eta}$ without otherwise changing the comb shape, consistent with the fact that loss commutes with the (linear) DP-MZM transformation when applied at the input or output of the device; a photon lost partway through the interferometer removes population uniformly from the sideband distribution and reduces the single-photon detection probability accordingly. We do not carry loss all the way to the density-matrix (mixed-state) description needed to capture loss-induced decoherence between branches, which we identify as an open extension in Section 8.

### 6.5 Stochastic impairments: RF phase noise, modulation-index jitter, and bias drift

Sections 6.1–6.3 treat bias error and I/Q imbalance as fixed (deterministic) mis-calibrations, which is the right picture for a systematic offset that can, in principle, be measured and corrected. In practice, DP-MZM transmitters are also subject to genuine stochastic noise: RF oscillator phase noise on each branch drive, shot-to-shot amplitude jitter in the RF driver electronics, and slow quadrature-bias drift from thermal and mechanical fluctuations acting on the static bias point. We model each as an independent, zero-mean Gaussian perturbation of the corresponding parameter in the general transfer function of Eq. (10); the complete Monte Carlo methodology, noise-model parameters, and random-seed/convergence details are given in Appendix C,

$$\theta_X \to \theta_X + \delta\theta_X, \quad m_X \to m^* + \delta m_X, \quad \phi_P \to \frac{\pi}{2} + \delta\phi_P, \qquad \delta \sim \mathcal{N}(0, \sigma^2) \qquad (14)$$

where $X \in \{I, Q\}$ labels the branch, $m^*$ is the optimum operating point (see Section 7), and $\sigma$ is the noise standard deviation (in degrees for the two phase-type noises, as a fraction of $m^*$ for the index jitter). RF phase noise and modulation-index jitter are applied independently to the I and Q branches (uncorrelated per-branch RF electronics), while bias drift acts as a single, device-wide random variable per realization ($\phi_P$ is one shared static bias point). For each noise type and each value of σ, we compute the results after $N = 500$ independent realizations $\delta_k$ and evaluate the qubit fidelity of Eq. (12) for each, reporting the Monte Carlo mean and standard deviation,

$$\langle F\rangle(\sigma) = \frac{1}{N}\sum_{k=1}^{N} F\left(H(q;\delta_k)\right), \qquad \delta_k \overset{\text{i.i.d.}}{\sim} \mathcal{N}(0,\sigma^2) \qquad (15)$$

Figure 10 shows $\langle F\rangle(\sigma) \pm 1\sigma$ for all three noise mechanisms. The most immediate observation is that the mean fidelity is comparatively insensitive to all three noise sources near σ = 0: $\langle F\rangle$ stays within 0.001 of the noise-free value $F^* \simeq 0.854$ up to $\sigma_\varphi \simeq 10°$ of RF phase noise, and degrades only gradually thereafter, reaching $\langle F\rangle \simeq 0.830$ at $\sigma_\varphi = 25°$. This robustness is a direct consequence of the operating point $m^*$ being, by construction (Section 7), a local maximum of the fidelity: the leading-order response to a zero-mean symmetric perturbation around a maximum is quadratic, so small stochastic noise costs far less mean fidelity than an equal-sized systematic (one-sided) offset of the kind studied in Section 6.1–6.3. Modulation-index jitter and bias drift show the same qualitative robustness, with $\langle F\rangle \simeq 0.843$ at $\sigma_m/m^* =$ 20% and $\langle F\rangle \simeq 0.843$ at $\sigma_{\Delta\phi_P} = 25°$, respectively.

The trial-to-trial spread tells a different, and practically important, story. At the largest noise levels shown, the standard deviation across realizations reaches ±0.104 for RF phase noise, compared with only ±0.045 for modulation-index jitter and ±0.021 for bias drift at their respective largest σ — i.e., RF phase noise produces roughly twice the realization-to-realization fidelity spread of index jitter, and five times that of bias drift, even though all three produce a similar mean degradation. This matters for any single-shot or few-shot quantum protocol (as opposed to a power- or ensemble-averaged classical measurement): a source with large realization-to-realization variance can produce individual low-fidelity events well below the reported mean, even when the mean itself looks acceptable. This identifies RF phase noise, rather than modulation-index jitter or bias drift, as the noise mechanism most warranting active suppression (e.g., a low-phase-noise RF source or phase-locked drive) in a DP-MZM-based single-photon or frequency-bin qubit source.

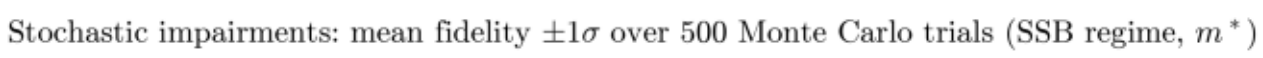

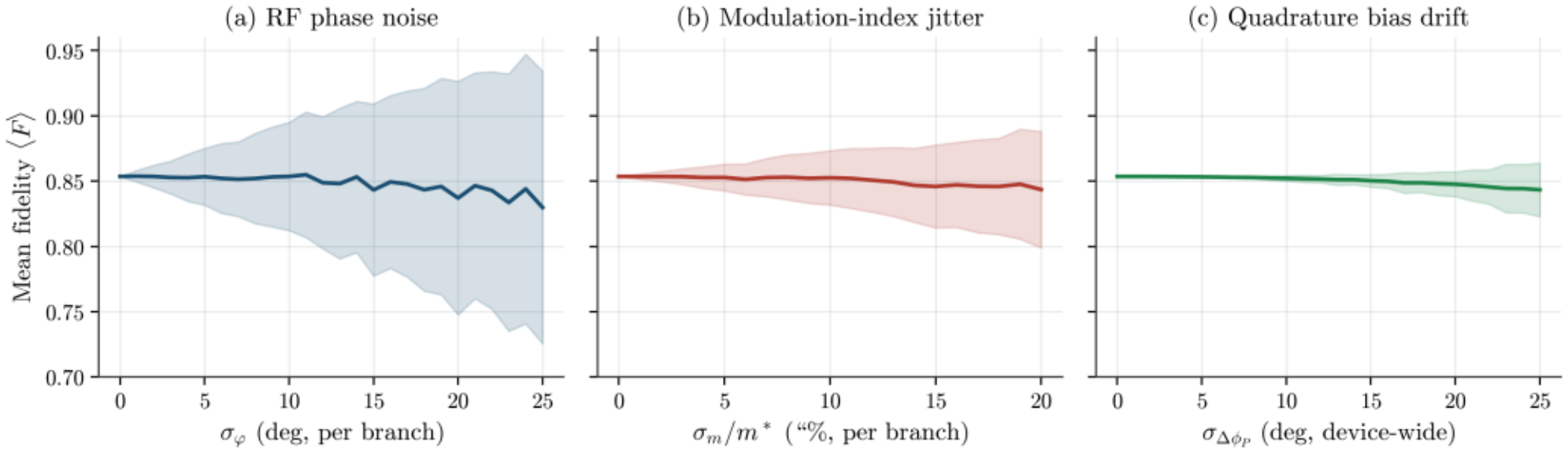

Fig. 10. Monte Carlo mean fidelity ⟨F⟩ (solid line) ± 1 standard deviation (shaded band) over 500 trials, at the (optimum) m* operating point, as a function of the noise standard deviation for (a) independent per-branch RF phase noise, (b) independent per-branch modulation-index jitter, and (c) device-wide quadrature-bias drift.

## 7. Application: Frequency-Bin Qubit State Preparation

As a concrete application connecting the model to the quantum-information scenarios outlined in Section 1 (refs. [7]–[10]), we use the DP-MZM transfer function of Eq. (10), at the RF-quadrature (single-sideband) operating point established in Sections 5–6, to design an operating point that prepares a frequency-bin qubit encoded in the carrier ($q = 0$) and first sideband ($q = +1$) frequency modes, using the fidelity metric of Eq. (12).

Scanning the modulation index at this operating point to maximize the fidelity of Eq. (12) gives $m^* \simeq 1.161$, at which the $q = 0$ and $q = 1$ populations are essentially balanced, as shown in Fig. 11 (48.8% and 48.8% of the total single-photon click probability, respectively, with the remaining ≈ 4.7% leaking into the neighboring $q = -1, 2, 3, \ldots$ bins). The resulting fidelity, however, saturates at $F^* \simeq 0.854$, not unity — and does not improve with further tuning of $m$. The reason is a fixed relative phase, not a population imbalance: at this operating point, $H(0)$ carries phase of +45° while $H(1)$ is exactly real (phase 0°), so the realized state is $\left(|q=0\rangle + e^{i45^\circ}|q=1\rangle\right)/\sqrt{2}$ up to normalization, rather than the equal-phase target of Eq. (12); this fixed 45° offset is what limits the overlap to $\cos^2(45°/2) \simeq 0.854$ even with perfectly balanced populations. Because this phase offset is deterministic and set entirely by the RF-quadrature operating condition $\theta_Q = \theta_I - \pi/2$, it is not, strictly speaking, a fidelity-degrading impairment: a receiver or downstream protocol that defines its logical qubit basis as $\left\{|q=0\rangle,\ e^{i45^\circ}|q=1\rangle\right\}$ rather than the naively equal-phase $\{|q=0\rangle,\ |q=1\rangle\}$ recovers unit fidelity trivially, since the offset is fixed and known rather than random. We retain the naive equal-phase target in Eq. (12) throughout this section because it cleanly isolates the additional fidelity loss caused by the impairments of Section 6 from this fixed, correctable basis offset.

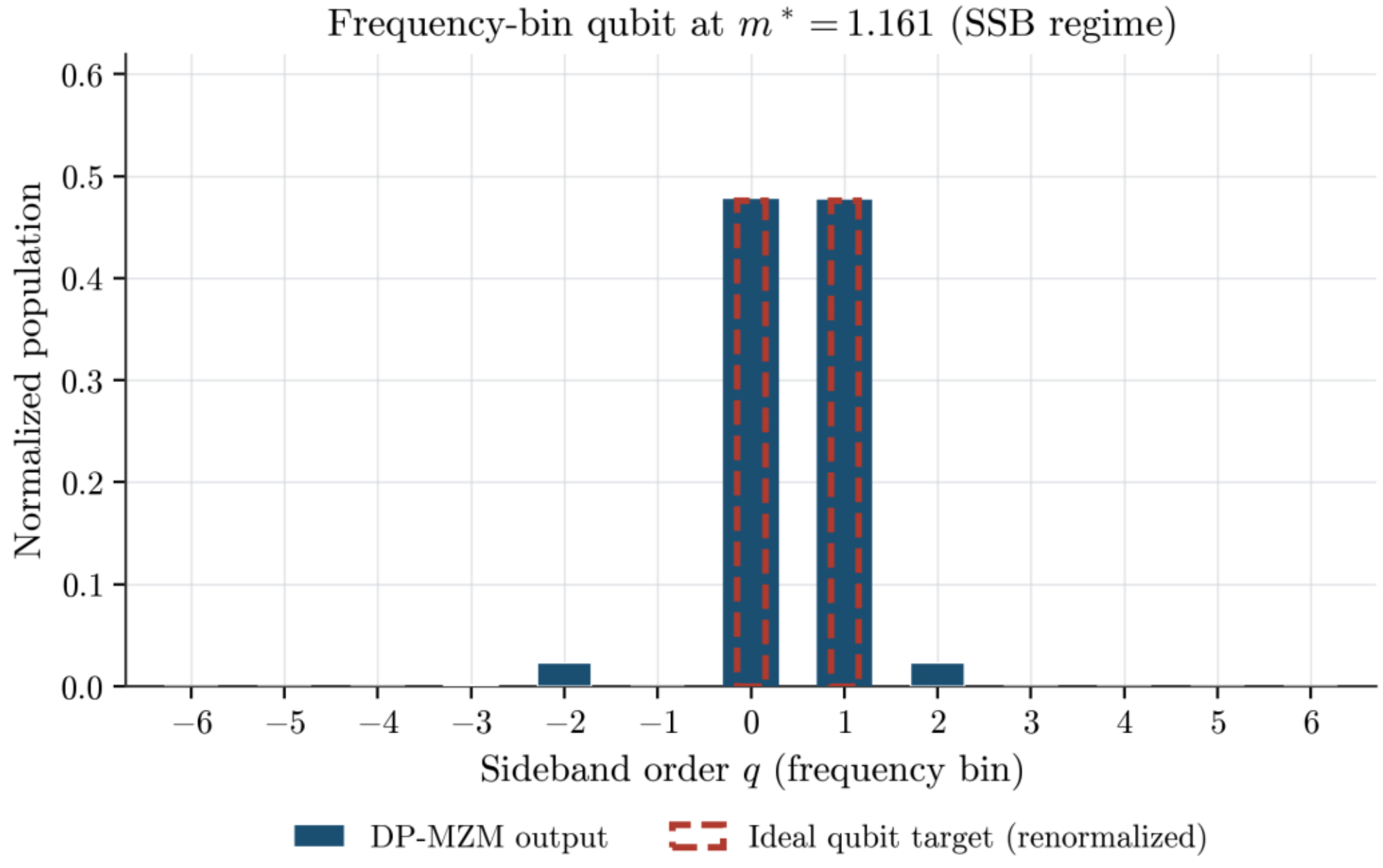

Fig. 11. DP-MZM output sideband populations at the frequency-bin qubit operating point m* ≃ 1.161, RF-quadrature regime (blue bars), compared with the population-balanced target at q = 0, 1 (red dashed outline, renormalized to the q∈{0,1} subspace population). The fidelity to the naive equal-phase target (Eq. (12)) is capped at F* ≃ 0.854 by the fixed 45° relative phase discussed in the text, not by population imbalance.

Combining the impairment mechanisms accounted for in Section 6, Fig. 12 maps the fidelity of the realized frequency-bin qubit to the naive equal-phase target as a joint function of quadrature-bias error and I/Q modulation-index imbalance, at the operating point $m^*$.

Consistent with the fixed 45° phase-offset ceiling identified above, the fidelity varies only modestly across the parameter ranges shown (F ≃ 0.807–0.875), remaining close to the F* ≃ 0.854 ceiling; bias error and I/Q imbalance perturb the realized state around that ceiling rather than being the dominant source of fidelity loss, which is instead the fixed phase convention discussed above. Adopting the phase-corrected target basis noted above would shift this entire map upward by essentially a constant offset, isolating the impairment-driven fidelity loss directly.

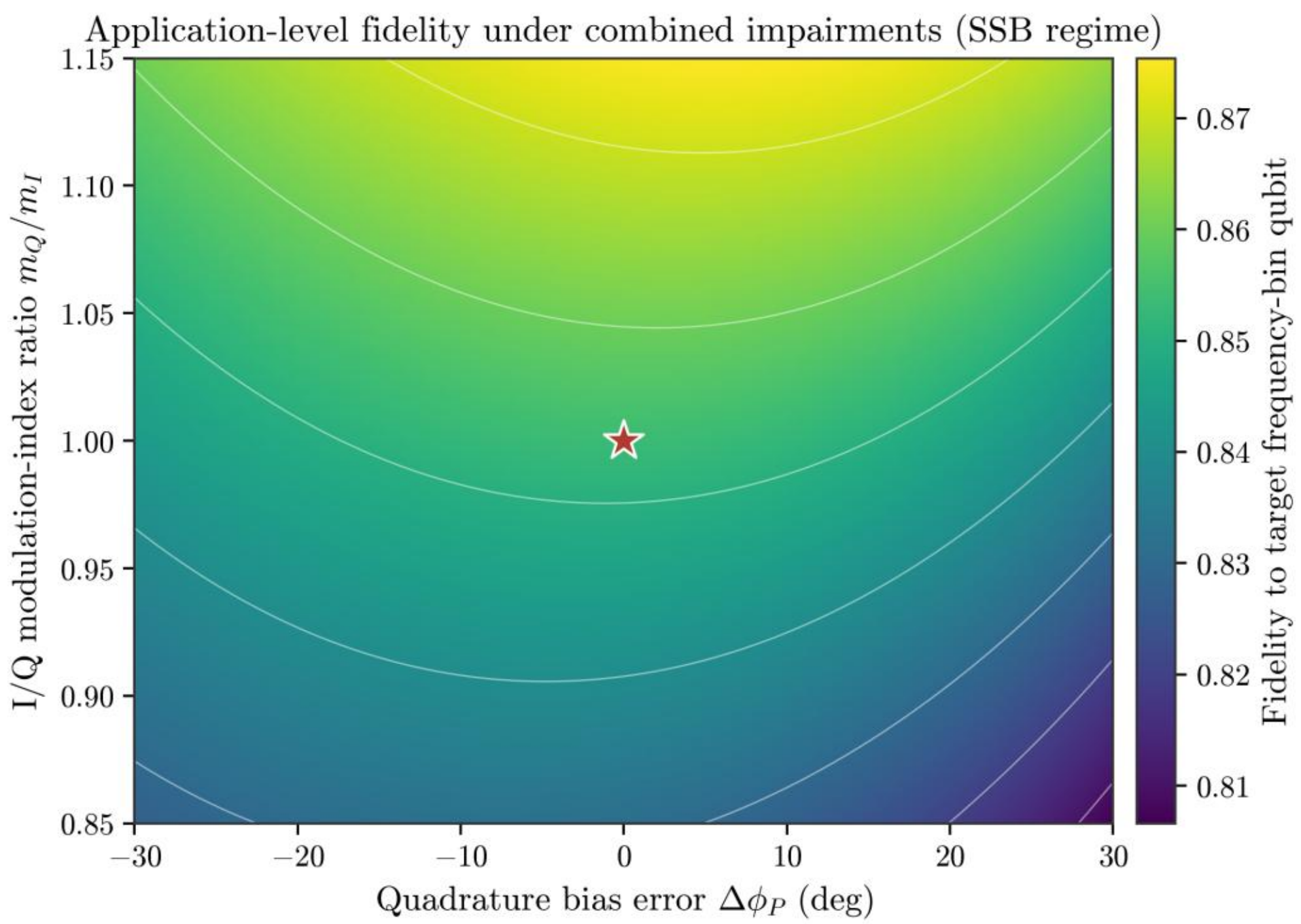

Fig. 12. Frequency-bin qubit fidelity (Eq. (12), naive equal-phase target) as a joint function of quadrature bias error and I/Q modulation-index ratio, at the m* operating point of Fig. 11. The red star marks the ideal operating point (Δϕ_P = 0, m_Q/m_I = 1).

The same mechanism generalizes naturally to a frequency-bin qutrit (populating q = −1, 0, +1, in the spirit of the electro-optic frequency tritters of ref. [7]) by relaxing the two-bin fidelity target of Eq. (12) to a three-bin overlap with an appropriately defined target phase convention; we leave a full treatment of higher-dimensional frequency-bin encodings, and an explicit connection to QKD decoy-state source characterization along the lines of ref. [10], as directions for future work (Section 8).

## 8. Conclusions

We have extended the quantum scattering formalism of electro-optic modulation to the general, unrestricted dual-parallel Mach–Zehnder modulator (DP-MZM) architecture, deriving the complete single-photon and coherent-state transformations for arbitrary beamsplitter ratios, independent branch modulation indices and bias phases, and independent in-phase/quadrature RF drive frequencies. The ideal, balanced, push-pull special case of Sections 3.3 and 4.3 is recovered exactly as a limit of this general model.

Using the general model, we identified two physically distinct operating regimes: the symmetric push-pull point of Sections 3.3/4.3, at which sideband amplitudes are always balanced between $+q$ and $-q$ and no image-sideband suppression is possible in principle; and an RF-quadrature (single-sideband) operating point at which the device suppresses one half of the sideband spectrum and which is the physically correct baseline for image-suppression and frequency-selective applications. Within the latter regime, we numerically quantified the sensitivity of the device to quadrature-bias error, I/Q modulation-index imbalance, finite beamsplitter extinction ratio, and, via Monte Carlo averaging, stochastic RF phase noise, modulation-index jitter, and quadrature-bias drift. The stochastic and deterministic pictures give complementary information: mean fidelity is comparatively robust to all three noise sources near the optimal operating point, since that point is a local fidelity maximum, but RF phase noise produces a substantially larger realization-to-realization spread than the other two mechanisms, identifying it as the impairment most warranting active mitigation in practice. We then applied the model to design and evaluate a DP-MZM-based frequency-bin qubit source, finding that its fidelity to a naive equal-phase target state saturates at $F^* \simeq 0.854$ — not because of residual population imbalance, which can be tuned away almost exactly, but because of a fixed, deterministic 45° relative phase between the carrier and first-sideband bins that is intrinsic to the RF-quadrature operating condition. This ceiling is not a fundamental limitation: it is removable by adopting a phase-corrected logical qubit basis, a distinction that only becomes visible once the general, regime-aware model developed here is used instead of the single, symmetric operating point alone.

Throughout this paper, the radio-frequency (RF) signal driving each phase modulator has been treated as a classical, externally prescribed field: the modulation index $m$ and bias phase $\theta_b$ enter the model as c-number (classical, non-operator) parameters, and it is precisely this classical-drive assumption that makes the Jacobi–Anger-type expansion of Eqs. (2)–(4) exactly solvable in closed form. A natural and increasingly relevant extension is the regime in which the modulating signal is itself a quantum microwave field rather than a classical drive — the situation encountered in microwave-to-optical quantum transduction, where the electro-optic (Pockels) effect used here is applied specifically to convert quantum states between superconducting microwave circuits and telecom-band photons [12]. Quantizing the RF drive replaces the c-number $m$ in the phase-modulator operator with a genuine microwave field operator, turning the interaction from one that is linear in the optical mode operators alone into a tripartite interaction coupling the optical input, the optical sidebands, and the microwave mode. This interaction is no longer, in general, exactly solvable by the closed-form Bessel-function approach used throughout Sections 3–4: the standard route in the transduction literature is instead to linearize the interaction around a strong classical optical or microwave pump (recovering an effective beamsplitter- or two-mode-squeezing-type Hamiltonian solvable by input–output theory), or to truncate the microwave mode to a small photon-number subspace and treat the resulting finite-dimensional problem numerically. Extending the DP-MZM quantum scattering formalism developed here to a quantized microwave drive — and, in particular, connecting the resulting single- and few-microwave-photon transfer functions to the frequency-comb structure and impairment analysis of Sections 5–6 — is a future direction of research linking this work to the broader effort of building hybrid microwave-optical quantum networks.

Overall, the model, results and examples developed here establish the general DP-MZM quantum scattering model as a practical design and analysis tool for photonic quantum state engineering. Additional natural directions for future work include a full open-quantum-system (Lindblad master-equation) treatment unifying insertion loss and RF phase noise into a single, jointly-derived decoherence framework (rather than the Monte Carlo and amplitude-rescaling treatments used here), generalization of the single-photon result to arbitrary Fock and entangled input states, the quantized-RF-drive extension discussed above, and an explicit, numerically evaluated connection between the frequency-bin application of Section 7 and quantum key distribution decoy-state source characterization.

Funding. Ministerio de Ciencia y Universidades, Plan Complementario de Comunicación Cuántica (QUANTUMABLE-1, QUANTUMABLE-2); HUB de Comunicaciones Cuánticas.

Disclosures. The authors declare no conflicts of interest.

Data availability. Data underlying the results presented in this paper are not publicly available at this time but may be obtained from the authors upon reasonable request.

## Appendix A: Detailed Stage-by-Stage Derivation — Single-Photon Input

This appendix reproduces, in full, the stage-by-stage propagation of the single-photon creation operator through the DP-MZM that underlies the general result of Eq. (5), following the device order of Fig. 1: input beamsplitter, input internal MZM beamsplitters, internal phase modulators, output internal combiners and quadrature bias, and output beamsplitter. Equation numbers in this appendix are prefixed 'A' and are referenced from Section 3.

### *A.1 Input beamsplitter action*

The input mode $\hat{a}_{no}^\dagger \otimes 1$ (single photon at the physical input port, vacuum at the fictitious port) is split by $\hat{S}_{BS,i}$ into the in-phase and quadrature arms:

$$\begin{pmatrix} \hat{a}_i^{(I)\dagger} \\ \hat{a}_i^{(Q)\dagger} \end{pmatrix} = \hat{S}_{BS,i}\left(\hat{a}_{n_o}^\dagger \otimes \mathbb{1}\right)\hat{S}_{BS,i}^\dagger = \begin{pmatrix} t_i\,\hat{a}_0^\dagger \\ -r_i\,\hat{a}_0^\dagger \end{pmatrix} \tag{A1}$$

### *A.2 Input internal MZM beamsplitters action*

Each arm mode is further split by its internal input beamsplitter, $\hat{S}_{BS,1}^{(I)}$ and $\hat{S}_{BS,1}^{(Q)}$, into the upper (+) and lower (−) internal paths that will carry the two push-pull phase modulators:

$$\begin{pmatrix} \hat{a}_+^{(I)\dagger} \\ \hat{a}_-^{(I)\dagger} \end{pmatrix} = \hat{S}_{BS,1}^{(I)}\left(\hat{a}_i^{(I)\dagger} \otimes \mathbb{1}\right)\hat{S}_{BS,1}^{(I)\dagger} = \begin{pmatrix} t_{I,1} t_i\,\hat{a}_0^\dagger \\ -r_{I,1} t_i\,\hat{a}_0^\dagger \end{pmatrix}, \qquad \begin{pmatrix} \hat{a}_+^{(Q)\dagger} \\ \hat{a}_-^{(Q)\dagger} \end{pmatrix} = \hat{S}_{BS,1}^{(Q)}\left(\hat{a}_i^{(Q)\dagger} \otimes \mathbb{1}\right)\hat{S}_{BS,1}^{(Q)\dagger} =$$

$$= \begin{pmatrix} -t_{Q,1} r_i\,\hat{a}_0^\dagger \\ r_{Q,1} r_i\,\hat{a}_0^\dagger \end{pmatrix} \tag{A2}$$

### *A.3 Internal MZM phase-modulator actions*

Each of the four resulting modes then passes through its respective phase modulator. Since the modulator operators act linearly on the mode they are fed, the constant amplitude prefactors from Eqs. (A1)–(A2) simply factor out of the (in general nonlinear, frequency-comb-generating) modulator transformation of Eq. (3):

$$\hat{S}_{PM,+}^{(I)}\hat{a}_+^{(I)\dagger}\hat{S}_{PM,+}^{(I)\dagger} = t_{I,1} t_i\,\hat{S}_{PM,+}^{(I)}\hat{a}_0^\dagger\hat{S}_{PM,+}^{(I)\dagger}, \quad \hat{S}_{PM,-}^{(I)}\hat{a}_-^{(I)\dagger}\hat{S}_{PM,-}^{(I)\dagger} = -r_{I,1} t_i\,\hat{S}_{PM,-}^{(I)}\hat{a}_0^\dagger\hat{S}_{PM,-}^{(I)\dagger} \tag{A3}$$

$$\hat{S}_{PM,+}^{(Q)}\hat{a}_+^{(Q)\dagger}\hat{S}_{PM,+}^{(Q)\dagger} = -t_{Q,1} r_i\,\hat{S}_{PM,+}^{(Q)}\hat{a}_0^\dagger\hat{S}_{PM,+}^{(Q)\dagger}, \quad \hat{S}_{PM,-}^{(Q)}\hat{a}_-^{(Q)\dagger}\hat{S}_{PM,-}^{(Q)\dagger} = r_{Q,1} r_i\,\hat{S}_{PM,-}^{(Q)}\hat{a}_0^\dagger\hat{S}_{PM,-}^{(Q)\dagger} \tag{A4}$$

### *A.4 Output internal combiners and quadrature bias*

The upper and lower paths of each internal MZI are recombined by the internal output beamsplitters $\hat{S}_{BS,2}^{(I,Q)}$; the quadrature-arm result additionally acquires the static bias phase $e^{i\phi_P}$ at this stage:

$$\hat{a}_o^{(I)\dagger} = \hat{S}_{BS,2}^{(I)}\begin{pmatrix} \hat{S}_{PM,+}^{(I)}\hat{a}_+^{(I)\dagger}\hat{S}_{PM,+}^{(I)\dagger} \\ \hat{S}_{PM,-}^{(I)}\hat{a}_-^{(I)\dagger}\hat{S}_{PM,-}^{(I)\dagger} \end{pmatrix}^T \hat{S}_{BS,2}^{(I)\dagger} = t_{I,2}\,\hat{S}_{PM,+}^{(I)}\hat{a}_+^{(I)\dagger}\hat{S}_{PM,+}^{(I)\dagger} + r_{I,2}\,\hat{S}_{PM,-}^{(I)}\hat{a}_-^{(I)\dagger}\hat{S}_{PM,-}^{(I)\dagger} \tag{A5}$$

$$\hat{a}_o^{(Q)\dagger} = \hat{S}_{BS,2}^{(Q)}\begin{pmatrix} \hat{S}_{PM,+}^{(Q)}\hat{a}_+^{(Q)\dagger}\hat{S}_{PM,+}^{(Q)\dagger} \\ \hat{S}_{PM,-}^{(Q)}\hat{a}_-^{(Q)\dagger}\hat{S}_{PM,-}^{(Q)\dagger} \end{pmatrix}^T \hat{S}_{BS,2}^{(Q)\dagger} = e^{i\phi_P} t_{Q,2}\,\hat{S}_{PM,+}^{(Q)}\hat{a}_+^{(Q)\dagger}\hat{S}_{PM,+}^{(Q)\dagger} +$$

$$e^{i\phi_P} r_{Q,2}\,\hat{S}_{PM,-}^{(Q)}\hat{a}_-^{(Q)\dagger}\hat{S}_{PM,-}^{(Q)\dagger} \tag{A6}$$

### *A.5 Output beamsplitter action*

Finally, the in-phase and quadrature arm outputs are combined by the output beamsplitter $\hat{S}_{BS,o}$. Its two output ports are the physical output mode (top component) and the fictitious radiated mode (bottom component), which is discarded:

$$\hat{S}_{BS,o}\begin{pmatrix}\hat{a}_o^{(I)\dagger}\\ \hat{a}_o^{(Q)\dagger}\end{pmatrix}\hat{S}_{BS,o}^{\dagger}=\begin{pmatrix}t_o & r_o\\ -r_o & t_o\end{pmatrix}\begin{pmatrix}\hat{a}_o^{(I)\dagger}\\ \hat{a}_o^{(Q)\dagger}\end{pmatrix}=\begin{pmatrix}t_o\hat{a}_o^{(I)\dagger}+r_o\hat{a}_o^{(Q)\dagger}\\ t_o\hat{a}_o^{(Q)\dagger}-r_o\hat{a}_o^{(I)\dagger}\end{pmatrix} \quad \text{(A7)}$$

Substituting Eqs. (A3)–(A6) into the physical (top) component of Eq. (A7) and collecting terms by branch reproduces the general single-photon result quoted as Eq. (5) in Section 3.2.

### *A.6 Extension to powers of the creation operator*

Because $\hat{S}_{DP}$ is unitary ($\hat{S}_{DP}^{\dagger}\hat{S}_{DP}$ = 1), the transformation of Eq. (5) extends immediately to arbitrary powers of the creation operator, and its Hermitian adjoint gives the corresponding annihilation-operator transformation used in Section 4.1:

$$\hat{S}_{DP}\,\hat{a}_{n_o}^{\dagger k}\,\hat{S}_{DP}^{\dagger}=\left(\hat{S}_{DP}\,\hat{a}_{n_o}^{\dagger}\,\hat{S}_{DP}^{\dagger}\right)^{k},\qquad \left(\hat{S}_{DP}\,\hat{a}_{n_o}^{\dagger}\,\hat{S}_{DP}^{\dagger}\right)^{\dagger}=\hat{S}_{DP}\,\hat{a}_{n_o}\,\hat{S}_{DP}^{\dagger} \quad \text{(A8)}$$

This extension is the technical bridge used in Appendix B: since a coherent state's displacement operator is built from an exponential series in $\hat{a}_n^{\dagger}$ and $\hat{a}_n$, Eq. (A8) guarantees that every term of that series transforms consistently under $\hat{S}_{DP}$, which is what allows the single-photon result of this appendix to be applied to the coherent-state case via the Baker–Campbell–Hausdorff argument of Appendix B.

### Appendix B: Detailed Stage-by-Stage Derivation — Coherent-State Input

This appendix repeats the derivation of Appendix A for a coherent-state input $\hat{D}_{no}(\alpha_{no}) \otimes 1$, underlying the general result of Eq. (8). Equation numbers are prefixed 'B' and are referenced from Section 4.

#### *B.1 Action of a unitary operator on a displacement operator: general lemma*

For any unitary $\hat{S}$ and any single-mode displacement operator $\hat{D}_n(\alpha_n) = \exp(\alpha_n\hat{a}_n^\dagger - \alpha_n^*\hat{a}_n)$, inserting $\hat{S}^\dagger\hat{S} = 1$ between each exponential factor and using $\hat{S}\, e^{\hat{X}} \hat{S}^\dagger = e^{\hat{S}\hat{X}\hat{S}^\dagger}$ for any operator $\hat{X}$ gives

$$\hat{S}\,\hat{D}_n(\alpha_n)\,\hat{S}^\dagger = \exp\left(-\frac{|\alpha_n|^2}{2}\right)\left[\hat{S}\,e^{\alpha_n\hat{a}_n^\dagger}\hat{S}^\dagger\right]\left[\hat{S}\,e^{-\alpha_n^*\hat{a}_n}\hat{S}^\dagger\right] = \exp\left(\alpha_n\,\hat{S}\hat{a}_n^\dagger\hat{S}^\dagger - \alpha_n^*\,\hat{S}\hat{a}_n\hat{S}^\dagger\right) \quad \text{(B1)}$$

i.e., the displacement operator transforms into the exponential of the transformed field operators — which is again a (generally multi-mode) displacement-type generator, since $\hat{S}\hat{a}_n^\dagger\hat{S}^\dagger$ is itself a linear combination of output creation operators by Eq. (A8).

#### *B.2 Reduction to a product of single-mode displacement operators*

Substituting the single-photon branch result (Appendix A) for $\hat{S}\hat{a}_{no}^\dagger\hat{S}^\dagger$ into Eq. (B1) gives a sum of output-mode creation and annihilation operators in the exponent. Because operators belonging to different output frequency bins commute, the Baker–Campbell–Hausdorff theorem, $\exp(\hat{A}+\hat{B}) = \exp([\hat{A},\hat{B}]/2)\exp(\hat{A})\exp(\hat{B})$, factorizes this joint exponential into a product of independent single-mode displacement operators, one per sideband — illustrated here for the upper in-phase branch modulator:

$$\hat{S}_{PM,+}^{(I)}\hat{D}_{n_o}(\alpha_{n_o})\hat{S}_{PM,+}^{(I)\dagger} \overset{\text{BCH}}{=} \prod_{q=1}^{\infty}\exp\left(\alpha_{qN-r}^{(I,+)}\hat{a}_{qN-r}^\dagger - \alpha_{qN-r}^{(I,+)*}\hat{a}_{qN-r}\right) =$$

$$= \prod_{q=1}^{\infty}\hat{D}_{qN-r}\left(\alpha_{qN-r}^{(I,+)}\right), \quad \alpha_{qN-r}^{(I,+)} = \alpha_{n_o}\,C_{q,+}^{(I)}\left(q_o^{(I)}\right) \quad \text{(B2)}$$

with identical structure for the lower in-phase and both quadrature-branch modulators. This product-of-displacement-operators structure is preserved through every subsequent linear stage of the device (beamsplitters, combiners, bias phase), since each such stage only reshuffles and re-weights which mode each displacement acts on, without mixing the exponents of different output bins.

#### *B.3 Input beamsplitter action*

$$\begin{pmatrix}\hat{D}_{n_o}^{(I)}\left(\beta_{n_o}^{(I)}\right)\\ \hat{D}_{n_o}^{(Q)}\left(\beta_{n_o}^{(Q)}\right)\end{pmatrix} = \hat{S}_{BS,i}\left(\hat{D}_{n_o}(\alpha_{n_o})\otimes\mathbb{1}\right)\hat{S}_{BS,i}^\dagger = \begin{pmatrix}\hat{D}_{n_o}\left(t_i\alpha_{n_o}\right)\\ \hat{D}_{n_o}\left(-r_i\alpha_{n_o}\right)\end{pmatrix} \quad \text{(B3)}$$

#### *B.4 Input internal MZM beamsplitters action*

$$\begin{pmatrix}\hat{D}_{n_o,+}^{(I)}\\ \hat{D}_{n_o,-}^{(I)}\end{pmatrix} = \begin{pmatrix}\hat{D}_{n_o}\left(t_{I,1}t_i\alpha_{n_o}\right)\\ \hat{D}_{n_o}\left(-r_{I,1}t_i\alpha_{n_o}\right)\end{pmatrix}, \qquad \begin{pmatrix}\hat{D}_{n_o,+}^{(Q)}\\ \hat{D}_{n_o,-}^{(Q)}\end{pmatrix} = \begin{pmatrix}\hat{D}_{n_o}\left(-t_{Q,1}r_i\alpha_{n_o}\right)\\ \hat{D}_{n_o}\left(r_{Q,1}r_i\alpha_{n_o}\right)\end{pmatrix} \quad \text{(B4)}$$

#### *B.5 Internal MZM phase-modulator actions*

Applying the results of Eq. (B1)–(B2) to each of the four branch modulators, with the amplitude prefactors of Eq. (B4) carried through:

$$\hat{S}_{PM,+}^{(I)}\hat{D}_{n_o,+}^{(I)}\hat{S}_{PM,+}^{(I)\dagger} = \prod_{q=1}^{\infty}\hat{D}_{qN-r}\left(t_{I,1}t_i\,\alpha_{n_o}\,C_{q,+}^{(I)}\left(q_o^{(I)}\right)\right), \qquad \hat{S}_{PM,-}^{(I)}\hat{D}_{n_o,-}^{(I)}\hat{S}_{PM,-}^{(I)\dagger} = \prod_{q=1}^{\infty}\hat{D}_{qN-r}\left(-r_{I,1}t_i\,\alpha_{n_o}\,C_{q,-}^{(I)}\left(q_o^{(I)}\right)\right) \tag{B5}$$

$$\hat{S}_{PM,+}^{(Q)}\hat{D}_{n_o,+}^{(Q)}\hat{S}_{PM,+}^{(Q)\dagger} = \prod_{q=1}^{\infty}\hat{D}_{qN-r}\left(-t_{Q,1}r_i\,\alpha_{n_o}\,C_{q,+}^{(Q)}\left(q_o^{(Q)}\right)\right), \qquad \hat{S}_{PM,-}^{(Q)}\hat{D}_{n_o,-}^{(Q)}\hat{S}_{PM,-}^{(Q)\dagger} = \prod_{q=1}^{\infty}\hat{D}_{qN-r}\left(r_{Q,1}r_i\,\alpha_{n_o}\,C_{q,-}^{(Q)}\left(q_o^{(Q)}\right)\right) \tag{B6}$$

*B.6 Output internal combiners and quadrature bias*

Recombining the upper and lower paths of each internal MZI, and including the quadrature bias phase $e^{i\phi_P}$:

$$\prod_{q=1}^{\infty}\hat{D}_{qN-r}^{(I)}\left(t_{I,2}t_{I,1}t_i\alpha_{n_o}C_{q,+}^{(I)}\left(q_o^{(I)}\right)\right)\hat{D}_{qN-r}^{(I)}\left(-r_{I,2}r_{I,1}t_i\alpha_{n_o}C_{q,-}^{(I)}\left(q_o^{(I)}\right)\right) \tag{B7}$$

$$\prod_{q=1}^{\infty}\hat{D}_{qN-r}^{(Q)}\left(-e^{i\phi_P}t_{Q,2}t_{Q,1}r_i\alpha_{n_o}C_{q,+}^{(Q)}\left(q_o^{(Q)}\right)\right)\hat{D}_{qN-r}^{(Q)}\left(e^{i\phi_P}r_{Q,2}r_{Q,1}r_i\alpha_{n_o}C_{q,-}^{(Q)}\left(q_o^{(Q)}\right)\right) \tag{B8}$$

*B.7 Output beamsplitter action and complete transformation*

Combining the in-phase and quadrature arm results at the output beamsplitter $\hat{S}_{BS,o}$ and retaining only the physical output port (discarding the fictitious radiated mode, as in Appendix A.5) gives the complete coherent-state DP-MZM transformation quoted as Eq. (8) in Section 4.2:

$$\hat{S}_{DP}\left(\hat{D}_{n_o}\left(\alpha_{n_o}\right)\otimes\mathbb{1}\right)\hat{S}_{DP}^{\dagger} \propto \prod_{q=1}^{\infty}\hat{D}_{qN_o-r_o}\left(\alpha_{n_o}\left[t_o t_{I,2}t_{I,1}t_i\,C_{q,+}^{(I)}\left(q_o^{(I)}\right) - t_o r_{I,2}r_{I,1}t_i\,C_{q,-}^{(I)}\left(q_o^{(I)}\right) - e^{i\phi_P}r_o t_{Q,2}t_{Q,1}r_i\,C_{q,+}^{(Q)}\left(q_o^{(Q)}\right) + e^{i\phi_P}r_o r_{Q,2}r_{Q,1}r_i\,C_{q,-}^{(Q)}\left(q_o^{(Q)}\right)\right]\right) \tag{B9}$$

As noted in Section 4.3, specializing Eq. (B9) to a common RF frequency, $\phi_P = \pi/2$, balanced beamsplitters, and push-pull operation reproduces Eq. (9).

## Appendix C: Monte Carlo Methodology

This appendix details the Monte Carlo procedure, noise-model parameters, and convergence checks underlying the stochastic-impairment results of Section 6.5 (Fig. 10).

### *C.1 Noise models and perturbed parameters*

All three noise mechanisms are evaluated at the fixed optimum operating point $m_I = m_Q = m^*$ $\simeq 1.161$ (Section 7), RF-quadrature bias $\theta_Q = \theta_I - \pi/2$, and ideal 3-dB balanced beamsplitters — i.e., every parameter other than the one being perturbed is held at its Section 7 nominal value. The three models, already summarized in Eq. (14), are applied as follows:

- RF phase noise: independent per-branch perturbations $\theta_I \rightarrow \theta_I + \delta\theta_I$ and $\theta_Q \rightarrow \theta_Q + \delta\theta_Q$, with $\delta\theta_I$, $\delta\theta_Q$ drawn independently from $\mathcal{N}(0,\sigma_\varphi^2)$ for each Monte Carlo trial, swept over $\sigma_\varphi \in \{0, 1, 2, \ldots, 25\}°$ (26 points).
- Modulation-index jitter: independent per-branch perturbations $m_I \rightarrow m^* + \delta m_I$ and $m_Q \rightarrow m^* + \delta m_Q$, with $\delta m_I$, $\delta m_Q$ drawn independently from $\mathcal{N}(0, \sigma_m^2)$, where $\sigma_m = (\sigma_m/m^*) \cdot m^*$ is parametrized as a percentage of m*, swept over $\sigma_m/m^* \in \{0,1,2,\ldots,20\}\%$ (21 points).
- Quadrature-bias drift: a single, device-wide perturbation $\phi_P \rightarrow \pi/2 + \delta\phi_P$ per trial (not independent per branch, since the static bias phase shifter of Fig. 1 is a single physical element), with $\delta\phi_P$ drawn from $\mathcal{N}\left(0, \sigma_{\Delta\phi_P}^2\right)$, swept over $\sigma_{\Delta\phi_P} \in$ $\{0,1,2,\ldots,25\}°$ (26 points).

### *C.2 Simulation procedure*

For each noise type and each value of σ on its respective grid, N = 500 independent noise realizations $\delta_k$ (k = 1,…,N) are drawn using a fixed pseudo-random generator seed for reproducibility. For each realization, the perturbed transfer function $H(q;\delta_k)$ is evaluated over the sideband range q ∈ {−6,…,6} using Eq. (10) with the perturbed parameter(s) substituted as in Appendix C.1, and the two-bin qubit fidelity $F(H(q;\delta_k))$ is computed via Eq. (12) restricted to the q ∈ {0,1} subspace. The Monte Carlo mean $\langle F\rangle(\sigma)$ and standard deviation across the N realizations are then formed as in Eq. (15) and plotted as the solid line and shaded ±1σ band, respectively, in each panel of Fig. 10.

### *C.3 Convergence*

With N = 500 trials per point, the standard error of the reported mean, $\mathrm{SE} = \sigma_F/\sqrt{N}$, is small relative to the effects discussed in Section 6.5: at the largest noise levels simulated, SE ≃ 0.0047 for RF phase noise ($\sigma_F \simeq 0.104$ at $\sigma_\varphi = 25°$), SE ≃ 0.0020 for modulation-index jitter ($\sigma_F \simeq 0.045$ at $\sigma_m/m^* = 20\%$), and SE ≃ 0.0009 for bias drift ($\sigma_F \simeq 0.021$ at $\sigma_{\Delta\phi_P} = 25°$) — in every case at least an order of magnitude smaller than the mean-fidelity variation being reported, confirming that N = 500 trials is sufficient for the mean curves of Fig. 10 to be stable to the precision quoted in the text. As a direct check, we repeated the worst-case point (RF phase noise, $\sigma_\varphi = 25°$) with $N = 1000$ trials at three independent random seeds; the resulting mean fidelities agreed with the N = 500 estimate to within 0.002, consistent with the SE estimate above.